\documentclass[runningheads]{llncs}

\usepackage{newunicodechar}
\usepackage{mathtools}
\usepackage{semantic}
\usepackage{qtree}
\usepackage{fontspec}
\usepackage{mathbbol}

\usepackage{caption}
\usepackage{subcaption}
\usepackage{proof}
\usepackage{xcolor}
\usepackage{amsmath}
\usepackage{amsfonts}
\usepackage[ruled]{algorithm2e}
\usepackage{hyperref}
\usepackage{url}
\usepackage{placeins}

\newcommand{\term}[2][]	{\ifthenelse{\equal{#1}{}}{\index{#2}}{\index{#1}}\emph{#2}}
\newcommand{\heading}[1] {\noindent {\bf #1}}

\newcommand{\code}[1]{\texttt{#1}}

\newcommand{\BD}{\begin{definition}}
\newcommand{\ED}{\end{definition}}

\newcommand{\idx}[2][]	{\ifthenelse{\equal{#1}{}}{\index{#2}#2}{\index{#1}#2}}

\newcommand{\eaci}[1]{\edef\tmp{\noexpand\index{\ac{#1}}} \emph{\tmp}}

\newcommand{\defvoc}[2]{\expandafter\def\csname #1\endcsname{#2} {\newacro{#1}{#2}}}

\newcommand{\vf}[1]{\def\tmp{\index{\csname #1\endcsname@{\acf{#1}}}}\tmp\emph{\acf{#1}}}
\newcommand{\vs}[1]{\def\tmp{\index{\csname #1\endcsname@{\acf{#1}}}}\tmp\acs{#1}}
\newcommand{\vl}[1]{\def\tmp{\index{\csname #1\endcsname@{\acl{#1}}}}\tmp\acl{#1}}

\SetKw{Throw}{throw}
\newcommand{\lean}{Lean}
\newcommand{\mlib}{Mathlib}

\newcommand{\derivat}{\code{HasDerivAt}}
\newcommand{\logth}{\code{HasDerivAt.log}}
\newcommand{\compth}{\code{HasDerivAt.comp}}
\newcommand{\diffcmd}{\code{differentiate}}
\newcommand{\cderiv}{\code{CDeriv}}

\usepackage{listings}

\usepackage{color}
\definecolor{keywordcolor}{rgb}{0.7, 0.1, 0.1}   
\definecolor{tacticcolor}{rgb}{0.0, 0.1, 0.6}    
\definecolor{commentcolor}{rgb}{0.4, 0.4, 0.4}   
\definecolor{symbolcolor}{rgb}{0.0, 0.1, 0.6}    
\definecolor{sortcolor}{rgb}{0.1, 0.5, 0.1}      
\definecolor{attributecolor}{rgb}{0.7, 0.1, 0.1} 

\newunicodechar{→}{\ensuremath{\to}}\newunicodechar{→}{\ensuremath{\to}}
\newunicodechar{ℝ}{\ensuremath{\mathbb{R}}}
\newunicodechar{𝕜}{\ensuremath{\mathbb{k}}}
\newunicodechar{𝓝}{\ensuremath{\mathcal{N}}}
\newunicodechar{∀}{\ensuremath{\forall}}
\newunicodechar{∃}{\ensuremath{\exists}}
\newunicodechar{ℤ}{\ensuremath{\mathbb{Z}}}
\newunicodechar{λ}{\ensuremath{\lambda}}
\newunicodechar{α}{\ensuremath{\alpha}}
\newunicodechar{β}{\ensuremath{\beta}}
\newunicodechar{τ}{\ensuremath{\tau}}
\newunicodechar{∘}{\ensuremath{\circ}}
\newunicodechar{≠}{\ensuremath{\neq}}
\newunicodechar{↦}{\ensuremath{\mapsto}}
\newunicodechar{∧}{\ensuremath{\wedge}}
\newunicodechar{₁}{\ensuremath{_1}}
\newunicodechar{₂}{\ensuremath{_2}}
\newunicodechar{₁}{\ensuremath{_1}}
\newunicodechar{₃}{\ensuremath{_3}}
\newunicodechar{¹}{\ensuremath{^1}}
\newunicodechar{⁻}{\ensuremath{^-}}

\title{Automatic Certified Differentiation in Lean}

\author{Ramy Shahin\orcidID{0000-0001-8724-3934}}
\institute{Qualgebra\\
\email{ramy@qualgebra.com}} 

\authorrunning{R. Shahin}

\titlerunning{Automatic Certified Differentiation in Lean}

\begin{document}

\maketitle

\begin{abstract}

Automatic differentiation (the automatic computation of numerical function derivatives) has many applications, especially in scientific computing and machine learning. 
Several approaches to automate the calculation of derivatives have been presented. Common challenges facing those approaches include 
(1) having to define derivative rules for different classes of functions, 
(2) having to manipulate program symbols to understand the semantics of the function being differentiated, and construct the derivative function, 
and (3), more importantly, ensuring that the calculated derivative is correct.

In this paper, we present an approach for automatic \emph{certified} differentiation using the Lean proof assistant. 
By certified, we mean a correctness certificate is computed together with the derivative function and a domain on which the derivative is guaranteed to be valid. 
Our approach leverages existing derivative theorems included in the Mathlib Lean library instead of having to define our own rules. 
We also utilize Lean metaprogramming to manipulate the symbolic structure of the input function, and to construct the derivative function.
The correctness certificate is computed step-wise as we apply Mathlib theorems.
To make the search efficient, we index the derivative theorems by the function they differentiate.
We evaluate our approach on 500 randomly generated functions: with the index, all of them are differentiated, with a median time of 0.18\,s per function, 8.6 times faster than without the index.


\end{abstract}

\section{Introduction}
\label{sec:intro}
\vspace{-0.1in}
\newcommand{\logfx}{\log(3/x)}
\newcommand{\drv}[1]{\frac{d #1}{dx}}

Given a mathematical function, computing its derivative has many practical applications.
This includes applications in scientific computing, machine learning, simulation, graphics, and computational finance.
The simplest approach is for a programmer to manually analyze the input function and write a program implementing the derivative function, following the differentiation rules in calculus textbooks.
However, since those rules have been well studied and understood for centuries, automating this process has been an active area of research since the 1960s~\cite{Wengert:1964}.

Approaches to computing derivatives automatically are usually classified into \emph{numerical}, \emph{algorithmic}, and \emph{symbolic differentiation}~\cite{Tolsma:1998,Baydin:2018}.
Numerical differentiation treats $f$ as a black box, and approximates its derivative at a given value $x$ by the rate of change of $f$ between $x$ and neighboring values.
Algorithmic differentiation evaluates the program computing $f$ at a given value $x$, and propagates derivative values through each elementary operation using the chain rule. It computes the exact value of the derivative at $x$, but not an expression for the derivative function.

On the other hand, symbolic differentiation is a white-box approach. 
It works directly on the syntactic expression structure of the function body without having to compute any function values. This is based on the fact that derivative rules are \emph{compositional}. 
For example, the chain rule is used to compute the derivative of a composition of two functions (e.g., $g(h(x))$). 
Following the syntactic tree structure of the body of function $f$, the derivative $\drv{f}$ is composed using primitive derivative theorems.

\begin{figure}[t]
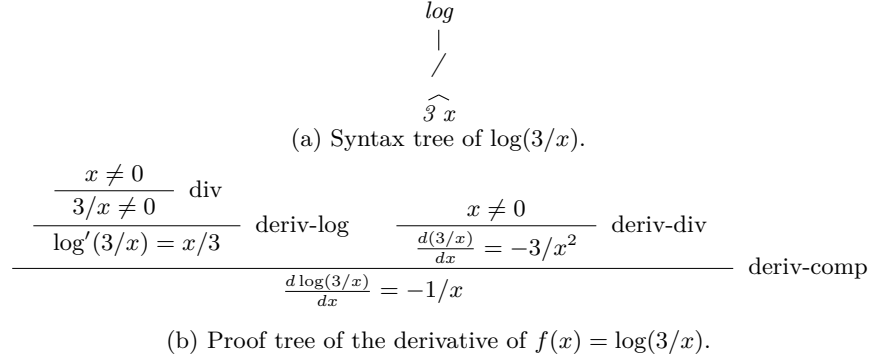

    \begin{subfigure}[]{\textwidth}
        \Tree[.\textit{log} 
            [.\textit{/} 
               \textit{3}
               \textit{x}
            ]
        ]
\vspace{-0.1in}
        \caption{Syntax tree of $\log(3/x)$.}
        \label{fig:tree}
    \end{subfigure}
    \vfill
    \vspace{-0.1in}
    \begin{subfigure}{\textwidth}

\[
    \inference 
        {\inference{\inference{x \neq 0}{3/x \neq 0}[div]}{\log'(3/x) = x/3}[deriv-log]
        &
        \inference{x \neq 0}{\drv{(3/x)} = -3/x^{2}}[deriv-div]
        }
        {\drv{\logfx} = -1/x}
        [deriv-comp]
\]
\vspace{-0.1in}
    \caption{Proof tree of the derivative of $f(x) = \log(3/x)$.}
    \label{fig:derivation}
    \end{subfigure}
    \caption{Syntax tree of the body of $f(x) = \log(3/x)$, and proof derivation of its derivative.}
    \label{fig:example}
    \vspace{-0.2in}
\end{figure}

For example, to differentiate the function $f = \logfx$, we follow its syntactic structure (Fig.~\ref{fig:tree}) top-down.
The root of the syntax tree is the unary function $\log$, and its only child is the binary function (operator) $/$.
This is a composition of two functions, and its derivative is computed using the chain rule. 
The two premises of the chain rule in this case are (1) the derivative of the outer function $\log$ applied to the inner function $3/x$, and (2) the derivative of the inner function $3/x$.

Using the logarithm derivative rule, the derivative of $\log$, evaluated at $3/x$, is $1/(3/x) = x/3$.
This rule only applies where the argument of $\log$ is non-zero, i.e., for $3/x \neq 0$, because the logarithm is not differentiable at zero.
In \mlib, division and the logarithm are total functions, with $x/0 = 0$ and $\log 0 = 0$, so side conditions like this one are about differentiability rather than definedness; since $3/0 = 0$, the condition $3/x \neq 0$ holds exactly when $x \neq 0$.

Similarly, using the quotient rule, the derivative of $3/x$ is $(0 \cdot x - 3 \cdot 1)/x^2 = -3/x^2$.
The quotient rule requires the denominator to be non-zero, so again the derivative is restricted to $x \neq 0$.
The derivative of $f$ now is the product of the two lower-level derivatives, according to the chain rule. 
This evaluates to $(x/3)(-3/x^2) = -1/x$.

This approach is completely based on formally proved theorems.
As a result, a correctness proof is (implicitly) constructed compositionally alongside the derivative function itself.
Fig.~\ref{fig:derivation} is the derivation tree of this proof.
This tree captures the exact same steps used to compute the derivative.
If this proof is made explicit, and returned by the automatic differentiation system together with the derivative function, then the derivative computation is \emph{certified}, i.e., it computes a value, together with a certificate proving the correctness of that value.

To be able to build such a system, three ingredients are needed: 
\begin{enumerate}
\item A formal foundation defining the notion of derivatives down to basic mathematical and logical principles, together with differentiation rules for different forms of mathematical expressions.
\item A way to validate the correctness of proofs built on top of that formal foundation.
\item The ability to manipulate the symbolic structure of mathematical functions, and to apply the right rule to the right kind of expression.
\end{enumerate}

The \lean~\cite{Moura:2021} interactive theorem prover together with its \mlib~\cite{Mathlib:2020} library provides all three ingredients.
(1) \mlib~is a large, community-maintained library of formalized mathematics, including analysis and calculus; its derivative theorems form exactly the required formal foundation.
(2) The \lean~kernel, which implements a variant of the Calculus of Inductive Constructions (CIC)~\cite{Bertot:2010}, only admits well-typed terms, so it acts as a proof checker for \mlib~and for proofs built on top of it.
(3) \lean~supports metaprogramming: programs that manipulate other programs, such as symbolic differentiation, and that extend the syntax of \lean~with new commands.

In this paper we present an automatic certified differentiation framework, leveraging the \lean~theorem prover, \mlib, and \lean~metaprogramming.
The correctness certificate returned together with the derivative function guarantees that the derivative is correct on the computed domain.
This is particularly relevant when used in developing safety-critical or mission-critical software systems.
Our contributions are:
\begin{itemize}
\item An algorithm for automatic certified differentiation that composes existing \mlib~derivative theorems following the structure of the input function, and returns the derivative, a domain on which it is valid, and a proof of its correctness, checked by the \lean~kernel (Sec.~\ref{sec:framework}).
\item An implementation of the algorithm as a \lean~command, including an index of derivative theorems by the function they differentiate, which makes the search efficient (Sec.~\ref{sec:implementation}).
\item An evaluation on 500 randomly generated functions, all of which are differentiated, with a median time of 0.18\,s per function, 8.6 times faster than without the index (Sec.~\ref{sec:eval}).
\end{itemize}

In particular, after covering some background on \mlib~derivative functions (Sec.~\ref{sec:background}), we present the automatic certified differentiation algorithm in Sec.~\ref{sec:framework}. 
We then cover the design of the library and how it integrates with \lean~in Sec.~\ref{sec:implementation}.
Evaluation results are then presented in Sec.~\ref{sec:eval}, and discussed in Sec.~\ref{sec:discussion}.
We then compare our approach to related work in Sec.~\ref{sec:related}, and finally conclude and discuss some future directions in Sec.~\ref{sec:conclusion}.
The library together with the evaluation driver and scripts is available in open source\footnote{\url{https://github.com/qualgebra/autodiff/tree/DATAMOD2026}}.
\section{Background}
\label{sec:background}

In this section we briefly introduce the concepts, constructs, definitions, and theorems of \lean~and \mlib~that we use in the rest of the paper.

\subsection{Differentiation in \mlib}
\label{sec:backgroundMathlib}

\begin{figure}[t]
\begin{minipage}[t]{0.48\textwidth}
    \vspace{0pt}
    \begin{subfigure}[]{\textwidth}
\begin{lstlisting}
def HasDerivAt
    { 𝕜 : Type u}
    [NontriviallyNormedField 𝕜]
    {F : Type v}
    [AddCommGroup F]
    [Module 𝕜 F]
    [TopologicalSpace F]
    [ContinuousSMul 𝕜 F]
(f : 𝕜 → F) (f' : F) (x : 𝕜): Prop
\end{lstlisting}
    \vspace{-0.1in}
    \caption{Definition of \derivat.}
    \label{fig:HasDerivAt}
    \end{subfigure}

    \begin{subfigure}[]{\textwidth}
\begin{lstlisting}
theorem HasDerivAt.log
    {f : ℝ → ℝ}
    {x f' : ℝ}
    (hf : HasDerivAt f f' x)
    (hx : ¬ (f x = 0)) :
HasDerivAt (fun (y : ℝ) => Real.log (f y))
    (f' / f x) x
\end{lstlisting}
    \vspace{-0.1in}
    \caption{Derivative of $\log$ composed with \code{f}.}
    \label{fig:log}
    \end{subfigure}
\end{minipage}\hfill
\begin{minipage}[t]{0.48\textwidth}
    \vspace{0pt}
    \begin{subfigure}[]{\textwidth}
\begin{lstlisting}
theorem HasDerivAt.comp
    { 𝕜 : Type u}
    [NontriviallyNormedField 𝕜]
    (x : 𝕜)
    { 𝕜' : Type u_1}
    [NontriviallyNormedField 𝕜']
    [NormedAlgebra 𝕜 𝕜']
    {h : 𝕜 → 𝕜'}
    {h₂ : 𝕜' → 𝕜'}
    {h' h₂' : 𝕜'}
    (hh₂ : HasDerivAt h₂ h₂' (h x))
    (hh : HasDerivAt h h' x):
HasDerivAt (h₂ ∘ h) (h₂' * h') x
\end{lstlisting}
    \vspace{-0.1in}
    \caption{Chain rule.}
    \label{fig:comp}
    \end{subfigure}
\end{minipage}
\vspace{-0.05in}
\caption{Documentation of some derivative definitions and theorems from \mlib.}
\label{fig:mathlib}
\vspace{-0.2in}
\end{figure}

\mlib~currently has two formalizations of derivatives: one-dimensional derivatives\footnote{Mathlib.Analysis.Calculus.Deriv}, and Fréchet derivatives\footnote{Mathlib.Analysis.Calculus.FDeriv}.
The framework presented in this paper only uses one-dimensional derivatives, restricting our scope to automatic differentiation of functions of a single variable.
Fig.~\ref{fig:mathlib} includes the documentation of the \mlib~definitions and theorems used in the examples in this paper.

The relation between a function $f$ and its derivative $f'$ at a given point $x$ is defined by the \derivat~predicate (Fig.~\ref{fig:HasDerivAt}).
The definition is not specific to functions on real numbers.
Instead, the domain of $f$ can be any nontrivially normed field $\mathbb{k}$. 
Its codomain can be any additive commutative group $F$.
Scalar multiplication between $\mathbb{k}$ and $F$ is also assumed to be jointly continuous in both of its arguments, and $F$ should have a topology defined on it.
All those constraints on $\mathbb{k}$ and $F$ are required by the underlying formalization of derivatives, and are encoded as type class instances.
\derivat~is a ternary predicate, relating a function \code{f} to its derivative \code{f'} at point \code{x}.

Derivative rules for different forms of mathematical expressions are defined in \mlib~as theorems in terms of \derivat.
For example, the logarithm derivative rule theorem is stated in Fig.~\ref{fig:log}. The theorem combines the logarithm derivative together with the chain rule, assuming that $\log$ is applied to a function $f$ of $x$, not $x$ itself.
The straightforward derivative of the logarithm of $x$ is an instance of this theorem, where $f$ becomes the identity function.
Unlike the generic definition of \derivat, this theorem is specific to $\mathbb{R}$, which satisfies all the type class constraints of \derivat.

Preconditions of the derivative rule are also included as premises in the theorem.
First, \code{hf} is a premise indicating that \code{f'} is the derivative of \code{f}.
Then \code{hx} is another premise restricting the argument of \code{log}, \code{f(x)} in this case, to non-zero values.
Such constraints on the variable in which the derivative is computed are essential for proof correctness.
For the automatic differentiation to generate a correctness certificate, it has to provide those input constraints as well.

Another example of a derivative rule theorem in \mlib~is the chain rule, or the derivative of function composition (Fig.~\ref{fig:comp}).
The structure of the theorem is similar to that of the logarithm derivative theorem.
This one keeps the domain and codomain types generic though, with the codomain of $h$ being the domain of $h_2$.
Proofs of all of those derivative theorems are included in \mlib, and are checked by the \lean~kernel when \mlib~is compiled; proofs built on top of them are checked by the kernel as well, so they cannot be used to construct higher-level proofs that are unsound.

\subsection{\lean~Metaprogramming}

\lean~includes a powerful set of metaprogramming constructs and libraries that enable users to write proof tactics, syntax extensions, and environment extensions, all in the \lean~language itself.
Proof tactics are commonly used to decompose proof goals into simpler sub-goals, and prove each of them individually.
The higher-level proof is automatically composed of the smaller proofs as a part of this process.
\lean~comes with an extensive set of standard general-purpose tactics.
Users can still write their own custom tactics using metaprogramming~\cite{Baanen:2024,Paulino:2025}.

When trying to prove a theorem in \lean, each proof goal becomes the type of a \lean~\emph{metavariable}.
Writing a proof is the process of assigning syntactic terms of the appropriate types to those metavariables.
The \lean~kernel does the type checking, admitting the proof only if all metavariables are assigned terms of the right types.
A tactic typically tries to automate much of the process of finding sound metavariable assignments.

A metavariable can also be used to encode missing information; such metavariables are sometimes called \emph{holes} in the proof.
When \lean~elaborates a syntactic term into a kernel expression, its unification algorithm can find type-sound assignments for those holes.
For example, the goal of this project is to find the derivative \code{f'} of a function \code{f} at point \code{x}, a correctness proof and a domain predicate.
The correctness theorem has to be of the form \derivat~\code{f f' x}.
However, \code{f'} is an unknown, so it cannot be explicitly included in the theorem until it is found.
As discussed in Sec.~\ref{sec:framework} and Sec.~\ref{sec:implementation}, a metavariable is used as a placeholder for \code{f'}, and the \lean~unification algorithm fills it in.
\section{Automatic Certified Differentiation Algorithm}
\label{sec:framework}

\LinesNumbered
\DontPrintSemicolon

\SetKwProg{KwProc}{procedure}{}{}
\SetKw{KwEProc}{end procedure}

\begin{algorithm}[t]
    \caption{Derivative Search Algorithm}
    \label{alg:diff}
    \KwProc{findDeriv}{}
    \KwIn{history, domain}
    \KwOut{domain}
    goal $\gets$ getCurrentGoal

    \If{goal $\in$ history}{
        fail
    }
    \If{the function in goal does not depend on x}{
        apply the constant-function theorem to goal\;
        \Return domain
    }
    \For{t $\in$ lookup(theoremDB, goal)} {
        \If{t applies to goal}{
            sg $\gets$ getSubgoals

            \For{s $\in$ sg} {
                \eIf{s has no metavariables}{
                    domain $\gets$ domain $\cup$ \{s\}
                }
                {
                    domain $\gets$ findDeriv (history $\cup$ \{goal\}) domain
                }
            }
            \Return domain
        }
    }
    fail\;
    \KwEProc
\end{algorithm}

Given a function $f:\alpha \to \beta$ (where types $\alpha$ and $\beta$ have instances of the type classes discussed in Sec.~\ref{sec:backgroundMathlib}), the goal of the automatic certified differentiation algorithm is to find three artifacts:
\begin{enumerate}
\item The derivative $f': \alpha \to \beta$.
\item A predicate $a:\alpha \to \textit{Prop}$ specifying a domain on which $f'$ is guaranteed to be the derivative.
\item A proof of the derivative correctness theorem \\
$\forall (x:\alpha), a \ x \to \textit{\derivat} \ f \ (f' \ x) \ x$.
\end{enumerate}

Finding all three artifacts is broken down into two stages.
The first stage finds the derivative $f'$, collecting the conjuncts of the domain constraint and an incomplete proof along the way.
The second stage completes the proof by filling in its missing pieces with the corresponding domain constraint conjuncts.

\subsection{Derivative Search}

The derivative search algorithm initially collects all theorems of the form 
\[ (\derivat \ \_ \ \_ \ \_) \] 
in the current environment. 
This set of theorems is referred to as the theorem database (\emph{theoremDB}).
The proof goal is then set to $G: \forall x, ?a \ x \to \derivat~f \ (?d \ x) \ x$, where $?a$ and $?d$ are \lean~metavariables for the domain predicate and the derivative.

Algorithm~\ref{alg:diff} is then used to find a potentially incomplete proof of this goal, while finding an assignment for the derivative metavariable $?d$. The algorithm takes two arguments: \emph{history}, which is the set of previously seen goals on a single branch of a recursion tree, and \emph{domain}, which is the set of domain constraints collected so far. When the algorithm starts, the empty set is passed for both arguments.

The algorithm starts by focusing on the current \emph{goal} (line 2). If it is included in the history, the call fails (lines 3--5).
This is a safeguard against falling into infinite recursion due to sequences of theorem applications that loop back to a previously visited step in the proof.
If the function being differentiated does not depend on $x$, it is constant, and the theorem stating that the derivative of a constant function is zero is applied directly, with no domain constraints (lines 6--9).
Otherwise, the candidate theorems for the goal are tried one by one (lines 10--22); with the theorem index (Sec.~\ref{sec:indexing}), these are only the theorems that can match the outermost function of the goal.
When a theorem is applicable to the goal (i.e., the unification algorithm finds a value assignment that matches both the theorem and the goal), the theorem is applied, and any generated subgoals are collected (lines 11--12).

For each subgoal $s$, if $s$ has no metavariables, it is a side condition of the theorem, such as $x \neq 0$, and it is added to the \emph{domain} set (lines 14--15).
The subgoal $s$ is left as a hole in the global proof, to be filled in later.
Otherwise, $s$ is another \derivat~goal, whose derivative is still to be found, and \code{findDeriv} is called recursively on $s$, adding \emph{goal} to the history of goals (line 17).
If all subgoals are solved, the algorithm returns the collected domain (line 20).
If a recursive call fails, the application of the theorem is undone, and the next candidate theorem is tried.
If no candidate theorem leads to a solution, the call fails (line 23).

The goal state is assumed to be global, and updated whenever a theorem is applied successfully to a goal.
When the algorithm succeeds, the original goal $G$ is assigned a proof.
This proof is incomplete, because it contains holes for the domain constraints.
Metavariable $?d$ is also instantiated with a function satisfying the theorem $G$.

\subsection{Domain Completion}

In the second stage, the holes in the proof corresponding to domain constraints are filled in.
First, a domain predicate is constructed as a conjunction of the domain constraints returned by \code{findDeriv}, and assigned to the metavariable $?a$.
The proof $G$ has a hypothesis of type $?a \ x$, i.e., a proof of that conjunction.
For each of the holes in the proof, the conjunct corresponding to its type is extracted from this hypothesis, using the \emph{left} and \emph{right} eliminators of conjunctions, and assigned to the hole.

\section{Implementation}
\label{sec:implementation}
\vspace{-0.2in}

\begin{figure}[t]
\begin{minipage}[t]{0.46\textwidth}
    \centering
    \vspace{0pt}
    \includegraphics[width=\textwidth]{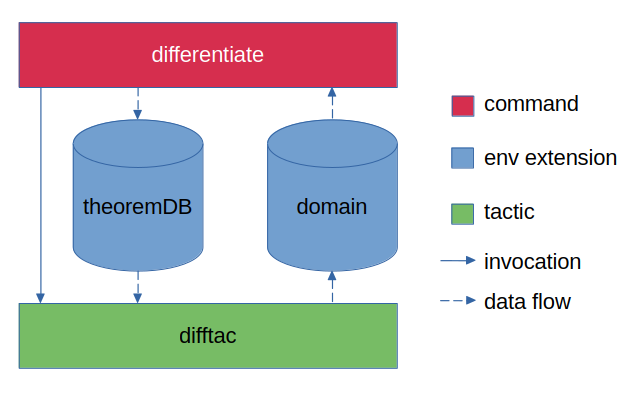}
    \vspace{-0.25in}
    \caption{Tool architecture.}
    \label{fig:arch}
\end{minipage}\hfill
\begin{minipage}[t]{0.5\textwidth}
    \vspace{0pt}
    \begin{lstlisting}
structure CDeriv {α β: Type}
  [NontriviallyNormedField α]
  [AddCommGroup β] [Module α β]
  [TopologicalSpace β] [ContinuousSMul α β]
  (f: α → β)
where
  f': α → β
  domain: α → Prop
  proof: ∀ (x:α), domain x →
    HasDerivAt f (f' x) x
    \end{lstlisting}
    \vspace{-0.2in}
    \captionsetup{justification=raggedright,singlelinecheck=false}\caption{The \cderiv~structure.}
    \label{fig:struct}
\end{minipage}
\vspace{-0.2in}
\end{figure}

\subsection{Design Decisions}
Fig.~\ref{fig:arch} is a high-level architecture of the tool implementation.
Differentiation inspects the syntactic structure of the function being differentiated, and it can fail, e.g., if the function is not differentiable, or if the packages containing the required theorems have not been imported.
A \lean~function cannot inspect the syntax of its argument, so we implement differentiation as a metaprogram: a \lean~syntactic extension command.
To differentiate a function \code{f}, the user issues the following command:
\begin{lstlisting}
let result := differentiate f
\end{lstlisting}
The result is a new definition \code{result} of type \cderiv~(Fig.~\ref{fig:struct}), a structure encapsulating the results of automatically differentiating a function \code{f} of type $\alpha \to \beta$, where $\alpha$ and $\beta$ satisfy the type class constraints of \derivat~(Sec.~\ref{sec:background}).
Its first field is the derivative \code{f'}, and its second field, \code{domain}, is a predicate on the function parameter type, defining a domain on which the derivative is valid.
The third field, \code{proof}, has the derivative correctness theorem as its type; its value is a proof that \code{f'} is the derivative of \code{f} at every point of \code{domain}.

The \diffcmd~command unfolds the body of function \code{f}, introduces its parameter, and runs the \code{difftac} tactic, which implements Alg.~\ref{alg:diff}.
The domain constraints found by the tactic are left as holes in the proof and recorded in a \lean~environment extension, through which they are communicated back to the command (outside the \code{TacticM} monad).
The command folds them into one conjunctive domain predicate, fills in each hole with the corresponding conjunct (Sec.~\ref{sec:framework}), and adds the resulting \cderiv~definition to the environment, where it is checked by the \lean~kernel.

When the \diffcmd~command is first used, it searches the current \lean~environment for theorems whose conclusion is a \derivat~statement (e.g., \logth~in Fig.~\ref{fig:log} and \compth~in Fig.~\ref{fig:comp}).
Those theorems are stored in \emph{theoremDB}, another \lean~environment extension, which serves as a derivative theorem database during derivative search.
It is assumed that the user will import the \mlib~packages relevant to the functions to be differentiated.
In addition to the \mlib~theorems, the database contains two theorems proved in our library from \mlib~theorems: a chain-rule form of the derivative of $\tan$, which \mlib~only states for $\tan$ itself, and a variant of \logth~with its premises reordered.
As a heuristic, the theorems are ordered in ascending order of their number of arguments.
To decide whether a theorem applies to a goal, the search uses \lean's unifier, which checks definitional (not just syntactic) equality; the same check is used to detect goals that are already in the history.

\subsection{Theorem Indexing}
\label{sec:indexing}
A naive search tries every theorem in the database on every goal, in order, until one applies.
Most of those attempts fail, and failures are expensive: to establish that a theorem does not apply, the unifier may unfold definitions deeply, e.g., \derivat~down to its underlying filter-based definition, or real-number functions such as \code{Real.cos} down to their definitions in terms of complex numbers.
We therefore index the database in two ways.
First, we exclude \emph{catch-all} theorems, whose conclusion is \derivat~\code{f f' x} for an arbitrary function \code{f} (e.g., \code{HasStrictDerivAt.hasDerivAt}, which derives \derivat~from a stronger notion of differentiability).
Such theorems unify with every goal, but their premises are not derivative goals that the search can discharge, so applying them only leads to failures deeper in the search.
Second, we index the remaining theorems by the head symbol of the function they differentiate, e.g., \code{Real.sin} for \code{HasDerivAt.sin}, or division for \code{HasDerivAt.div}.
For a goal whose function has head symbol $h$, only the theorems indexed by $h$ are tried, together with those whose head symbol is not fixed (e.g., the derivative of a constant function), preserving the order of the database.
Subgoals that are not \derivat~statements, e.g., about other notions of differentiability, cannot be solved by the search, and fail immediately.
Sec.~\ref{sec:eval} compares the performance of derivative search with and without the index.

\subsection{Example}

Assume we want to differentiate the function $f(x) = \log(3/x)$ of Sec.~\ref{sec:intro}, using the command:
\begin{lstlisting}
noncomputable def f := λ x:ℝ ↦ log (3 / x)
let d := differentiate f
\end{lstlisting}
This defines a \cderiv~structure \code{d}, whose derivative and domain are
\begin{lstlisting}
fun x => (0 * x - 3 * 1) / x ^ 2 / (3 / x)
fun x => x ≠ 0 ∧ 3 / x ≠ 0
\end{lstlisting}
The derivative is not simplified; on its domain, it is equal to $-1/x$.
The \code{difftac} tactic uses four derivative theorems:
\begin{enumerate}
    \item \code{HasDerivAt.log'}, the chain rule with \code{log} as the outer function. Its domain premise states that the argument of \code{log} has to be non-zero, i.e., $3/x \neq 0$.
    \item \code{HasDerivAt.fun\_div}, the quotient rule, to differentiate $3/x$. Its domain premise states that the denominator has to be non-zero, i.e., $x \neq 0$.
    \item \code{hasDerivAt\_const} and \code{hasDerivAt\_id'}, to differentiate the numerator $3$ and the denominator $x$. These theorems have no domain premises.
\end{enumerate}

\begin{figure}[t]
\begin{subfigure}[]{0.95\textwidth}
\begin{lstlisting}
fun (x: ℝ) (a: ?a x) =>
  HasDerivAt.log' ?h₁
    (HasDerivAt.fun_div (hasDerivAt_const x 3) (hasDerivAt_id' x) ?h₂)
\end{lstlisting}
\vspace{-0.1in}
\caption{Incomplete proof generated by the tactic, with holes for the domain constraints.}
    \label{fig:preproof}
\end{subfigure}
\begin{subfigure}[]{0.95\textwidth}
\begin{lstlisting}
fun (x: ℝ) (a: x ≠ 0 ∧ 3 / x ≠ 0) =>
  HasDerivAt.log' a.right
    (HasDerivAt.fun_div (hasDerivAt_const x 3) (hasDerivAt_id' x) a.left)
\end{lstlisting}
    \vspace{-0.1in}
    \caption{Complete proof, after the holes are filled in.}
    \label{fig:proof}
\end{subfigure}
\caption{Incomplete and complete correctness proofs of the derivative of $\log(3/x)$.}
\label{fig:proofs}
\vspace{-0.2in}
\end{figure}

Fig.~\ref{fig:preproof} is the proof generated by the tactic.
The type of the hypothesis \code{a} is the domain predicate, still the metavariable \code{?a} applied to \code{x}.
The hole \code{?h₁} is the domain premise of \code{HasDerivAt.log'}, and should be a proof of $3/x \neq 0$.
The hole \code{?h₂} is the domain premise of \code{HasDerivAt.fun\_div}, and should be a proof of $x \neq 0$.

The command collects the types of the holes, and folds them into the conjunctive domain predicate $x \neq 0 \wedge 3/x \neq 0$, which is assigned to \code{?a}.
Each hole is then assigned the conjunct of its type, extracted from \code{a}.
Fig.~\ref{fig:proof} is the complete proof for the derivative of function \code{f}: \code{?h₁} is replaced by \code{a.right}, a proof of $3/x \neq 0$, and \code{?h₂} by \code{a.left}, a proof of $x \neq 0$.

\section{Evaluation}
\label{sec:eval}

Beyond the simple functions used during development, we evaluate the tool on randomly generated functions that exercise many different derivative rules.

Function bodies are expression trees over integer literals between $-100$ and $100$, the variable raised to a natural number exponent between 1 and 10, the four arithmetic operators, trigonometric functions ($\sin$, $\cos$, and $\tan$), and natural logarithms.
Each inner node of an expression tree is one of those constructs, chosen with equal probability, and each leaf is either a literal or the variable.
Function bodies of expression tree depth up to 7 are generated from a fixed random seed.
We exclude functions that do not mention the variable, since constant functions are trivially differentiable, as well as functions consisting of a single literal or power of the variable, and duplicates, resulting in 500 distinct test functions of 4 to 67 nodes (median 19).
Because only a few \mlib~types implement instances of the type classes required by the \derivat~theorems, we limit our test set to $\mathbb{R}\to\mathbb{R}$ functions.
Division, $\log$, and $\tan$ were included particularly to test domain constraints.

The goals of the evaluation experiments are:
\begin{enumerate}
\item To ensure that the framework works as expected on a diverse set of mathematical functions.
\item To identify any limitations of the approach and the implementation. 
\item To measure the computational time of the automatic differentiation implementation, both on functions differentiable on the entire set of real numbers, and on functions whose differentiable domain is restricted.
\item To measure the performance improvement due to theorem indexing (Sec.~\ref{sec:indexing}), by running the same test cases with and without the index.
\end{enumerate}

This evaluation was performed on an 8-core AMD Ryzen 7 7730U processor, with 40\,GB RAM, running Linux kernel version 7.0.0-38-generic. We used Lean 4.33.1 with \mlib~v4.33.1.
Each test case was run in two configurations: with the theorem index (Sec.~\ref{sec:indexing}), and without it, i.e., trying all theorems of the database in order.
The theorem database is built once before each test case.
For each configuration, the \diffcmd~command was run five times, where we excluded the first run and averaged the running time of the other four runs.
We used the default \lean~heartbeat limit (200,000).

\begin{table}[t]
\centering
\caption{Time per differentiation in milliseconds, with and without the theorem index, over the 488 test cases that complete in both configurations, and the speedup due to indexing (time without the index divided by time with the index, per test case). The domain is the one computed by the tool, which is the same in both configurations.}
\label{tab:results}
\footnotesize
\setlength{\tabcolsep}{3pt}
\begin{tabular}{lr|rrr|rrr|rr}
 & & \multicolumn{3}{c|}{time with index (ms)} & \multicolumn{3}{c|}{time without index (ms)} & \multicolumn{2}{c}{speedup} \\
domain & $n$ & median & p95 & max & median & p95 & max & median & range \\
\hline
universal  &  66 &  67 & 216 &   323 &   439 & 1486 &  2401 &  5.9$\times$ & 1.6--8.9$\times$ \\
restricted & 422 & 202 & 417 & 1921 &  2871 & 7221 & 12275 & 13.8$\times$ & 1.2--71$\times$ \\
\hline
all        & 488 & 181 & 410 & 1921 &  2423 & 7045 & 12275 &  8.6$\times$ & 1.2--71$\times$ \\
\end{tabular}
\vspace{-0.1in}
\end{table}

\begin{figure}[t]
\includegraphics[width=\textwidth]{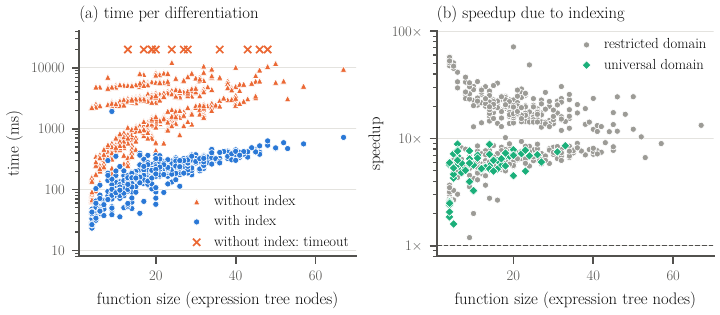}
\caption{(a) Time per differentiation in milliseconds against the size of the input function in expression tree nodes, with and without the theorem index (log scale). Crosses mark the 12 test cases that do not complete without the index. (b) Speedup due to indexing per test case, for functions whose computed domain is restricted or universal (log scale).}
\label{fig:graph}
\vspace{-0.2in}
\end{figure}

With the index, all 500 test cases ran to completion, generating the derivative of the input function, a domain on which the derivative is valid, and a proof that the generated derivative is correct, checked by the \lean~kernel.
Each of these proofs depends only on \lean's standard axioms (\code{propext}, \code{Classical.choice}, and \code{Quot.sound}); in particular, none of them uses \code{sorry}.
Without the index, 12 test cases did not complete because the \lean~heartbeat limit was reached; with the index, the same functions are differentiated in at most 0.6\,s.
For the 488 test cases that complete in both configurations, the computed domains are identical, and since both derivatives are certified on the same domain, they agree on it: the index changes which theorems are tried, and possibly the syntactic form of the derivative, but not the result of the search.
Table~\ref{tab:results} summarizes the running times, and Fig.~\ref{fig:graph} shows them per test case against the size of the input function in expression tree nodes.

With the index, the median time per differentiation is 181\,ms, 95\% of the test cases take less than 0.42\,s, and the slowest takes 1.9\,s.
Without the index, the median is 2.4\,s and the slowest successful test case takes 12.3\,s.
Indexing speeds up differentiation by a factor of 8.6 for the median test case, and the total time for all 488 test cases drops from 1333\,s to 98\,s.
The speedup grows with function size (Fig.~\ref{fig:graph}b), and is larger for functions with a restricted domain (median 13.8$\times$) than for functions with a universal domain (median 5.9$\times$).
Theorems with domain premises (e.g., for division, $\log$, and $\tan$) have more arguments than most other derivative theorems, so they come late in the database.
Without the index, the search reaches them only after trying, and failing to apply, most other theorems.

Functions with a restricted domain still take longer than those with a universal domain, also with the index (median 202\,ms vs.\ 67\,ms).
They tend to be larger (median 20 vs.\ 10.5 nodes) and to use more complex derivative theorems; in addition, each domain premise becomes a hole in the proof that has to be filled from the domain predicate.

We also checked the computed domains: every conjunct has the form $e \neq 0$ for some expression $e$, i.e., a non-zero denominator, argument of $\log$, or cosine of the argument of $\tan$, and all but one of the 433 restricted domains hold at some of 4000 points in $[-20, 20]$, evaluated with \mlib's conventions for division by zero and logarithms.
The exception, for $\log(\log((84 + (-84))(-98 - x^6)(29 - x)))$, requires a product with a zero factor to be non-zero, so its correctness theorem holds vacuously (Sec.~\ref{sec:discussion}).

\section{Discussion}
\label{sec:discussion}
\vspace{-0.1in}
\heading{Generality. }
Our evaluation (Sec.~\ref{sec:eval}) covered randomly generated real functions over a limited grammar, chosen for practical reasons: to evaluate the applicability and performance of the approach, whose soundness is guaranteed by the \lean~kernel.
The framework is not restricted to this grammar: it only needs derivative theorems for the building blocks of a function, and composition theorems such as the chain rule.
\mlib~has such theorems for many real functions outside our grammar, and they can be used as soon as their packages are imported; as \mlib~grows, it may also cover domains other than the real numbers, which the type class constraints of \derivat~already allow.
Functions with conditionals, such as piecewise functions, would need derivative theorems that hold on parts of the domain, which \derivat, being stated at a point, accommodates.
Functions of multiple variables and partial derivatives would require extending the \diffcmd~command, to compute a domain over all parameters and to specify the parameter to differentiate with respect to.

\heading{Soundness and Precision of Side Conditions. }
The correctness theorem $\forall x, \textit{domain}\ x \to \derivat\ f\ (f'\ x)\ x$ is a soundness guarantee: wherever the domain predicate holds, $f'$ is the derivative of $f$.
However, it does not guarantee that the domain predicate is precise.
Side conditions are collected from the premises of the theorems applied during derivative search, following the syntactic structure of $f$ rather than its semantics.
As a result, the computed domain is a sufficient, but not necessarily the weakest, precondition for differentiability, in the sense of Dijkstra's weakest preconditions~\cite{Dijkstra:1976}.

In the extreme case, the domain predicate can be unsatisfiable, making the correctness theorem vacuously true.
For example, since division in \mlib~is total, $f(x) = 1/(x-x)$ is constant, yet differentiating it syntactically yields the side condition $x - x \neq 0$, i.e., \textit{False}.
This also occurs in our evaluation, for one of the 500 test functions, where a factor of the argument of $\log$ is $84 + (-84)$ (Sec.~\ref{sec:eval}).
Such a certificate is useless, but it is never wrong: using the correctness proof at any particular point $x_0$ requires a proof of $\textit{domain}\ x_0$, so an unsatisfiable domain cannot be used to derive an incorrect derivative value.
The same limitation applies to side conditions generated by structural differentiation in other proof assistants (Sec.~\ref{sec:related}).

Two inexpensive mitigations are (1) simplifying the domain predicate and rejecting results whose domain reduces to \textit{False}, and (2) extending \cderiv~with a proof that the domain is non-empty, found by trying candidate witnesses.
A stronger guarantee is that the domain is a non-empty open set, so that the derivative is valid on a neighborhood of some point.
Proving that the domain is maximal, i.e., that $f$ is not differentiable outside of it, is outside the scope of this paper.

\heading{Performance and Further Optimizations. }
Profiling the search without the theorem index showed where its time goes.
For a typical function with a restricted domain, 97\% of the 1970 theorem applications attempted by the search failed, and those failed attempts took about 80\% of the running time.
Each attempt calls the unifier (\code{Lean.Meta.isDefEq}\footnote{\url{https://leanprover-community.github.io/mathlib4_docs/Lean/Meta/Basic.html\#Lean.Meta.isDefEq}}), which checks whether the function being differentiated is definitionally (not just syntactically) equal to the function \code{f} in the conclusion of a theorem.
Before giving up on a mismatch, the unifier unfolds definitions deeply, e.g., \derivat~down to its filter-based definition, and real-number functions down to their definitions in terms of complex numbers.
All test cases that time out without the index exhaust the heartbeat limit during unification.
The theorem index (Sec.~\ref{sec:indexing}) removes most of these attempts: for the same function, the number of attempted applications drops from 1970 to 79, and over the whole test set (Sec.~\ref{sec:eval}) differentiation becomes 8.6 times faster for the median test case, with no timeouts.

Further optimizations remain possible.
(1) Within the theorems indexed by a head symbol, the search still tries candidates in order with full unification. A discrimination tree, as used by \lean's simplifier, keyed on the structure of the whole function rather than its head symbol, would narrow the candidates further.
(2) Restricting unification to reducible definitions would make failed attempts cheap, but we found that some derivations rely on unfolding, e.g., of the instances of arithmetic operations on $\mathbb{R}$, so it would need a fallback to full unification.
(3) Preprocessing the input function into a syntactic normal form might allow syntactic instead of definitional matching altogether.

\vspace{0.1in}
\heading{Beyond Differentiation. }
One important reason why our framework works is the fact that derivatives are unique, i.e., a differentiable function has a unique derivative.
This guarantees the confluence of the stepwise construction of the proof tree of the derivative of a function.
As a result, we could start with a set of theorems (together with their proofs), and apply them compositionally to reach a computational result, together with a correctness certificate.
But is this unique to computing derivatives?

If we were to inspect the design of the framework again, it might be surprising that, with the exception of particularly storing theorems with a \derivat~conclusion in the database, it has nothing specific to computing derivatives.
The domain predicate of the derivative is a precondition of the compositional computation.
Other certified compositional computations can be inferred in a similar fashion.
For example, static program analyses based on Dijkstra's weakest precondition calculus~\cite{Dijkstra:1976} can be implemented similarly.
Here a precondition (ideally the weakest one) would be analogous to the derivative domain, while the result of the analysis would be analogous to the derivative.
Generalizing our framework to \emph{search} for computational results and their preconditions given compositional theorems, while stepwise constructing a correctness certificate as a part of the process, can be another future direction in which this work can evolve.
\section{Related Work}
\label{sec:related}
\vspace{-0.1in}

Automatic differentiation in its different forms (numerical, algorithmic, and symbolic) has been extensively surveyed by different research groups~\cite{Tolsma:1998,Hoffmann:2016,Baydin:2018}.
Other surveys focused on flaws in existing automatic differentiation systems~\cite{Johnson:2025}, and pitfalls developers typically overlook when developing new automatic differentiation approaches~\cite{Huckelheim:2024}. With so many issues found in existing systems, it is clear that generating a correctness certificate together with the derivative function is of real value.

Within \mlib~itself, the \code{fun\_prop} tactic proves properties such as differentiability by decomposing a function into elementary functions, and simplification lemmas about \code{deriv} can rewrite derivative expressions.
The SciLean scientific computing library~\cite{Skvrivan:2025} builds on \mlib's derivatives with the \code{fun\_trans} tactic, which computes derivatives by rewriting with differentiation rules registered with the tactic.
In both cases, side conditions such as $x \neq 0$ for division or $\log$ have to be discharged at a given point, e.g., from hypotheses supplied by the user.
In contrast, our approach infers those side conditions, and returns them as the domain on which the computed derivative is valid.
In previous work, we embedded Datalog in \lean, translating Datalog queries into theorems proved by backward chaining, and used it, among other examples, to compute derivatives from \mlib~theorems listed by the user~\cite{Shahin:2026}.
That example is restricted to functions differentiable everywhere, since it requires the derivative at every point; here, the applicable theorems are found automatically, and the domain on which the derivative is valid is computed.

Structural, theorem-driven differentiation has also been implemented in other proof assistants.
In HOL4, the \code{DIFF\_CONV} conversion~\cite{HOL4:Diff}, building on Harrison's formalization of real analysis in HOL~\cite{Harrison:1996}, derives a derivative theorem for a given term, carrying side conditions as hypotheses.
Isabelle/HOL provides the \code{derivative\_eq\_intros} rule collection, which, combined with the simplifier, computes derivatives of large expressions, leaving side conditions as residual subgoals~\cite{Paulson:2025}.
In Coq, the \code{auto\_derive} tactic of the Coquelicot library~\cite{Boldo:2015} proves the differentiability of an expression and computes its derivative, generating side goals for the conditions under which the derivative exists.
In a similar spirit, an automated continuity checker for real-valued functions was developed in PVS~\cite{Gottliebsen:2000}.
Our approach differs in three ways.
First, the result is a single self-contained \cderiv~object packaging the derivative, an explicit domain predicate, and a closed proof, rather than a theorem with hypotheses or residual goals left to the user.
Second, instead of relying on a fixed, curated rule set, derivative rules are discovered by searching the environment for \derivat~theorems, so new \mlib~theorems are picked up automatically once imported.
Third, differentiation is exposed as a top-level command producing a definition, rather than as a tactic used inside a proof.
A different use of theorem proving is to verify an automatic differentiation tool once and for all, as done in Coq for the O$\partial$yss\'ee tool on a subset of FORTRAN programs~\cite{Mayero:2002}.
In contrast, our approach does not verify the differentiation procedure itself; instead, each computed derivative comes with its own kernel-checked certificate.

Approaches other than theorem proving have also been pursued to improve the correctness of automatic differentiation systems.
At the formal semantics level, automatic differentiation formalization in the Turing-complete PCF core language has been studied~\cite{Mazza:2021}.
Correctness of typical approaches particularly in the case of non-differentiable functions was also studied~\cite{Lee:2020}.
Lighter-weight approaches were also proposed, e.g., static analysis to find bugs in automatic differentiation systems~\cite{Laurel:2023}.
Those approaches all start from basic principles, attempting to formally define the correctness of differentiation, and apply the definition to some classes of functions.
Our approach, on the other hand, starts with mathematically sound theorems whose proofs are checked by \lean, and builds a stepwise proof that follows the structure of the function being differentiated.

Using theorem provers to write certified programs or develop other software artifacts has been gradually growing in adoption and scope in the last 20 years. 
Early success stories included the CompCert certified C compiler project~\cite{Leroy:2009} and the seL4 certified microkernel~\cite{Klein:2009}.
More recent examples include formalizing authorization policies and access permissions~\cite{Cutler:2024}, formalizing change-impact assessment in product lines of assurance cases~\cite{Shahin:2021}, and sound compositional construction of assurance cases from building blocks~\cite{Viger:2023}.
\section{Conclusion}
\label{sec:conclusion}
\vspace{-0.1in}
In this paper we presented an approach to automatic function differentiation that generates a correctness certificate together with the derivative.
It also generates a predicate on the function input, a sufficient condition under which the derivative is valid.
This approach builds on \mlib's formalization of derivatives and differentiation rules.
Our evaluation of the approach shows that it handles relatively complex functions, including functions whose derivative is valid only on a restricted domain, and that indexing the derivative theorems by the function they differentiate makes the search 8.6 times faster for the median test case, and up to 71 times faster.

We believe this approach can be extended in several ways. Our implementation currently supports only first-order, one-dimensional derivatives, possibly with restricted domains.
One relatively straightforward extension is adding support for higher-order derivatives by repeatedly differentiating the derivative.
Supporting partial derivatives and multidimensional (Fréchet) derivatives, which \mlib~already formalizes, is another possible extension.
Finally, the computed domain is a sufficient condition for the derivative to be valid, but not necessarily the largest one; proving that it is non-empty, or open, would make the correctness certificate more informative (Sec.~\ref{sec:discussion}).

Beyond extending the approach itself, certified derivatives are ultimately meant to be used in software, such as safety-critical systems, where the computed derivative is evaluated using fixed-point or floating-point approximations of real numbers.
Formalizing the approximation bounds of those representations would extend the correctness guarantee from the derivative over the reals to its numerical implementation.

\begin{credits}
\subsubsection{\ackname}
We thank the anonymous reviewers for their insightful comments and suggestions.
\end{credits}

\bibliographystyle{splncs04}
\bibliography{pl,autodiff}

@misc{Baanen:2024,
  title={The hitchhiker’s guide to logical verification (2024 Edition)},
  author={Baanen, Anne and Bentkamp, Alexander and Blanchette, Jasmin and H{\"o}lzl, Johannes and Limperg, Jannis},
  url={https://lean-forward.github.io/hitchhikers-guide/2024/}
}

@article{Baydin:2018,
  title={{Automatic differentiation in machine learning: a survey}},
  author={Baydin, Atilim Gunes and Pearlmutter, Barak A and Radul, Alexey Andreyevich and Siskind, Jeffrey Mark},
  journal={Journal of machine learning research},
  volume={18},
  number={153},
  pages={1--43},
  year={2018}
}

@article{Cutler:2024,
author = {Cutler, Joseph W. and Disselkoen, Craig and Eline, Aaron and He, Shaobo and Headley, Kyle and Hicks, Michael and Hietala, Kesha and Ioannidis, Eleftherios and Kastner, John and Mamat, Anwar and McAdams, Darin and McCutchen, Matt and Rungta, Neha and Torlak, Emina and Wells, Andrew M.},
title = {Cedar: A New Language for Expressive, Fast, Safe, and Analyzable Authorization},
year = {2024},
issue_date = {April 2024},
publisher = {Association for Computing Machinery},
address = {New York, NY, USA},
volume = {8},
number = {OOPSLA1},
url = {https://doi.org/10.1145/3649835},
doi = {10.1145/3649835},
journal = {Proc. ACM Program. Lang.},
month = apr,
articleno = {118},
numpages = {28}
}

@article{Hoffmann:2016,
  title={{A hitchhiker’s guide to automatic differentiation}},
  author={Hoffmann, Philipp HW},
  journal={Numerical Algorithms},
  volume={72},
  number={3},
  pages={775--811},
  year={2016},
  publisher={Springer}
}

@article{Huckelheim:2024,
author = {Hückelheim, Jan and Menon, Harshitha and Moses, William and Christianson, Bruce and Hovland, Paul and Hascoët, Laurent},
title = {A taxonomy of automatic differentiation pitfalls},
journal = {WIREs Data Mining and Knowledge Discovery},
volume = {14},
number = {6},
pages = {e1555},
doi = {https://doi.org/10.1002/widm.1555},
url = {https://wires.onlinelibrary.wiley.com/doi/abs/10.1002/widm.1555},
eprint = {https://wires.onlinelibrary.wiley.com/doi/pdf/10.1002/widm.1555},
year = {2024}
}

@article{Johnson:2025,
title = {Software-based Automatic Differentiation is Flawed},
journal = {Journal of Computational Physics},
pages = {114319},
year = {2025},
issn = {0021-9991},
doi = {https://doi.org/10.1016/j.jcp.2025.114319},
url = {https://www.sciencedirect.com/science/article/pii/S002199912500600X},
author = {Daniel Johnson and Trevor Maxfield and Yongxu Jin and Ronald Fedkiw}
}

@inproceedings{Klein:2009,
        title={{seL4: Formal verification of an OS kernel}},
        author={Klein, Gerwin and Elphinstone, Kevin and Heiser, Gernot and Andronick, June and Cock, David and Derrin, Philip and Elkaduwe, Dhammika and Engelhardt, Kai and Kolanski, Rafal and Norrish, Michael and others},
        booktitle={Proceedings of the ACM SIGOPS 22nd Symposium on Operating Systems Principles},
        pages={207--220},
        year={2009}
}

@article{Laurel:2023,
author = {Laurel, Jacob and Qian, Siyuan Brant and Singh, Gagandeep and Misailovic, Sasa},
title = {Synthesizing Precise Static Analyzers for Automatic Differentiation},
year = {2023},
issue_date = {October 2023},
publisher = {Association for Computing Machinery},
address = {New York, NY, USA},
volume = {7},
number = {OOPSLA2},
url = {https://doi.org/10.1145/3622867},
doi = {10.1145/3622867},
journal = {Proc. ACM Program. Lang.},
month = oct,
articleno = {291},
numpages = {29}
}

@article{Lee:2020,
  title={On correctness of automatic differentiation for non-differentiable functions},
  author={Lee, Wonyeol and Yu, Hangyeol and Rival, Xavier and Yang, Hongseok},
  journal={Advances in neural information processing systems},
  volume={33},
  pages={6719--6730},
  year={2020}
}

@article{Leroy:2009,
        author = {Xavier Leroy},
        title = {{A Formally Verified Compiler Back-end}},
        journal = {Journal of Automated Reasoning},
        volume = 43,
        number = 4,
        pages = {363--446},
        year = 2009,
        url = {http://xavierleroy.org/publi/compcert-backend.pdf},
        urlpublisher = {http://dx.doi.org/10.1007/s10817-009-9155-4}, 
        hal = {http://hal.inria.fr/inria-00360768/},
        pubkind = {journal-int-mono}
}

@inproceedings{Mathlib:2020,
author = {The mathlib Community},
title = {The lean mathematical library},
year = {2020},
isbn = {9781450370974},
publisher = {Association for Computing Machinery},
address = {New York, NY, USA},
url = {https://doi.org/10.1145/3372885.3373824},
doi = {10.1145/3372885.3373824},
booktitle = {Proceedings of the 9th ACM SIGPLAN International Conference on Certified Programs and Proofs},
pages = {367–381},
numpages = {15},
location = {New Orleans, LA, USA},
series = {CPP 2020}
}

@article{Mazza:2021,
author = {Mazza, Damiano and Pagani, Michele},
title = {Automatic differentiation in PCF},
year = {2021},
issue_date = {January 2021},
publisher = {Association for Computing Machinery},
address = {New York, NY, USA},
volume = {5},
number = {POPL},
url = {https://doi.org/10.1145/3434309},
doi = {10.1145/3434309},
journal = {Proc. ACM Program. Lang.},
month = jan,
articleno = {28},
numpages = {27}
}

@inproceedings{Moura:2021,
  title = {The Lean 4 Theorem Prover and Programming Language},
  author = {Moura, Leonardo de and Ullrich, Sebastian},
  year = {2021},
  isbn = {978-3-030-79875-8},
  publisher = {Springer-Verlag},
  address = {Berlin, Heidelberg},
  url = {https://doi.org/10.1007/978-3-030-79876-5_37},
  doi = {10.1007/978-3-030-79876-5_37},
  booktitle = {Automated Deduction – CADE 28: 28th International Conference on Automated Deduction, Virtual Event, July 12–15, 2021, Proceedings},
  pages = {625–635},
  numpages = {11}
}

@misc{Paulino:2025,
  title={Metaprogramming in Lean 4},
  author={Paulino, Arthur and Testa, Damiano and Ayers, Edward and Karunus, Evgenia and B{\"o}ving, Henrik and Limperg, Jannis and Gadgil, Siddhartha and Bhat, Siddharth},
  url={https://leanprover-community.github.io/lean4-metaprogramming-book/}
}

@inproceedings{Shahin:2021,
  title={Towards certified analysis of software product line safety cases},
  author={Shahin, Ramy and Kokaly, Sahar and Chechik, Marsha},
  booktitle={International Conference on Computer Safety, Reliability, and Security},
  pages={130--145},
  year={2021},
  organization={Springer}
}

@misc{Skvrivan:2025,
  title={Scientific Computing in Lean},
  author={Sk{\v{r}}ivan, Tom{\'a}{\v{s}}},
  url={https://lecopivo.github.io/scientific-computing-lean/},
  year={2022}
}

@article{Tolsma:1998,
  title={On computational differentiation},
  author={Tolsma, John E and Barton, Paul I},
  journal={Computers \& chemical engineering},
  volume={22},
  number={4-5},
  pages={475--490},
  year={1998},
  publisher={Elsevier}
}

@article{Viger:2023,
  title={{The FOREMOST approach to building valid model-based safety arguments}},
  author={Viger, Torin and Murphy, Logan and Di Sandro, Alessio and Menghi, Claudio and Shahin, Ramy and Chechik, Marsha},
  journal={Software and Systems Modeling},
  volume={22},
  number={5},
  pages={1473--1494},
  year={2023},
  publisher={Springer}
}

@article{Wengert:1964,
  title={A simple automatic derivative evaluation program},
  author={Wengert, Robert Edwin},
  journal={Communications of the ACM},
  volume={7},
  number={8},
  pages={463--464},
  year={1964},
  publisher={ACM New York, NY, USA}
}

@article{Boldo:2015,
  title={{Coquelicot: A User-Friendly Library of Real Analysis for Coq}},
  author={Boldo, Sylvie and Lelay, Catherine and Melquiond, Guillaume},
  journal={Mathematics in Computer Science},
  volume={9},
  number={1},
  pages={41--62},
  year={2015},
  doi={10.1007/s11786-014-0181-1}
}

@inproceedings{Gottliebsen:2000,
  title={{Transcendental Functions and Continuity Checking in PVS}},
  author={Gottliebsen, Hanne},
  booktitle={Theorem Proving in Higher Order Logics (TPHOLs 2000)},
  series={LNCS},
  volume={1869},
  pages={197--214},
  publisher={Springer},
  year={2000},
  doi={10.1007/3-540-44659-1_13}
}

@techreport{Harrison:1996,
  title={{Theorem Proving with the Real Numbers}},
  author={Harrison, John},
  institution={University of Cambridge Computer Laboratory},
  number={UCAM-CL-TR-408},
  year={1996}
}

@misc{HOL4:Diff,
  title={{HOL4 Diff structure: DIFF\_CONV}},
  author={{The HOL4 Team}},
  howpublished={\url{https://hol-theorem-prover.org/docs/trindemossen-2/htmlsigs/Diff.html}},
  note={Accessed October 2026}
}

@inproceedings{Mayero:2002,
  title={{Using Theorem Proving for Numerical Analysis (Correctness Proof of an Automatic Differentiation Algorithm)}},
  author={Mayero, Micaela},
  booktitle={Theorem Proving in Higher Order Logics (TPHOLs 2002)},
  series={LNCS},
  volume={2410},
  pages={246--262},
  publisher={Springer},
  year={2002},
  doi={10.1007/3-540-45685-6_17}
}

@inproceedings{Paulson:2025,
  title={{Formalising New Mathematics in Isabelle: Diagonal Ramsey}},
  author={Paulson, Lawrence C.},
  booktitle={16th International Conference on Interactive Theorem Proving (ITP 2025)},
  series={LIPIcs},
  volume={352},
  pages={18:1--18:18},
  year={2025},
  doi={10.4230/LIPIcs.ITP.2025.18}
}

@inproceedings{Shahin:2026,
  title={{A Shallow Embedding of Datalog in Lean}},
  author={Shahin, Ramy},
  booktitle={Proceedings of the 19th ACM SIGPLAN International Conference on Software Language Engineering (SLE '26)},
  pages={108--117},
  publisher={ACM},
  year={2026},
  doi={10.1145/3806383.3815521}
}

@book{Bertot:2010,
	author = {Bertot, Yves and Castran, Pierre},
	title = {{Interactive Theorem Proving and Program Development: Coq'Art The Calculus of Inductive Constructions}},
	year = {2010},
	isbn = {3642058809, 9783642058806},
	edition = {1st},
	publisher = {Springer Publishing Company, Incorporated}
}

@book{Dijkstra:1976,
author = {Dijkstra, Edsger Wybe},
title = {A Discipline of Programming},
year = {1976},
isbn = {013215871X},
publisher = {Prentice Hall PTR},
address = {USA},
edition = {1st}
}

\end{document}